\documentclass[aip,jmp,amsmath,amssymb,reprint]{revtex4-1}

\usepackage{enumitem}
\usepackage{array}
\usepackage{float}
\usepackage{graphicx}
\usepackage{dcolumn}
\usepackage{bm}

\makeatletter

\newcommand{\Rmnum}[1]{\expandafter\@slowromancap\romannumeral #1@}
\makeatother

\begin{document}

\preprint{AIP/123-QED}

\title{A flowing plasma model to describe drift waves in a cylindrical helicon discharge}

\author{L. Chang}
\email{chang.lei@anu.edu.au.}
\author{M. J. Hole}
\author{C. S. Cormac}

\affiliation{ 
Plasma Research Laboratory, Research School of Physics and Engineering, \\The Australian National University, Canberra, ACT 0200, Australia
}%

\date{\today}

\begin{abstract}
A two-fluid model developed originally to describe wave oscillations in the vacuum arc centrifuge, a cylindrical, rapidly rotating, low temperature and confined plasma column, is applied to interpret plasma oscillations in a RF generated linear magnetised plasma (WOMBAT), with similar density and field strength. Compared to typical centrifuge plasmas, WOMBAT plasmas have slower normalised rotation frequency, lower temperature and lower axial velocity. Despite these differences, the two-fluid model provides a consistent description of the WOMBAT plasma configuration and yields qualitative agreement between measured and predicted wave oscillation frequencies with axial field strength. In addition, the radial profile of the density perturbation predicted by this model is consistent with the data. Parameter scans show that the dispersion curve is sensitive to the axial field strength and the electron temperature, and the dependence of oscillation frequency with electron temperature matches the experiment. These results consolidate earlier claims that the density and floating potential oscillations are a resistive drift mode, driven by the density gradient. To our knowledge, this is the first detailed physics model of flowing plasmas in the diffusion region away from the RF source. Possible extensions to the model, including temperature non-uniformity and magnetic field oscillations, are also discussed.
\end{abstract}

\maketitle

\section{introduction}
Oscillations have been observed in the electric probe signals of many laboratory plasmas, including theta pinches,\cite{Rostoker1961, Freidberg1978} magnetic mirrors,\cite{Perkins1963, Kuo1964, Hooper1983} Q machines,\cite{Motley1963} vacuum arc centrifuge \cite{Krishnan1981} and helicon plasmas.\cite{Boswell1983, Boswell1984, Degeling1999, Greiner1999, Sun2005} Theta pinches and magnetic mirrors behave vastly different from the latter because of their high beta and significant magnetic curvature, which result in highly different plasma and field geometry. In contrast, Q machine, plasma centrifuge and helicon plasmas have low beta and negligible magnetic curvature, and have similar plasma properties. We exploit this similarity by deploying a two-fluid collisional model, initially developed to describe vacuum arc centrifuge plasmas, to describe the configuration and oscillations in a helicon plasma. Our goal is to better describe a class of electrostatic oscillations observed, but not yet conclusively identified, in the diffusion region of helicon plasmas, away from the RF source. 

The helicon plasma we study is provided by the WOMBAT (Waves On Magnetised Beams And Turbulence)\cite{Boswell1987} device, which was designed to provide a plasma environment to study wave physics, including chaos, turbulence, wave saturation and the related driving mechanisms. WOMBAT has been used to study helicon waves,\cite{Ellingboe1996, Degeling1999, Corr2007} which have attracted great interest in the past decades due to their ability to produce high plasma densities. \cite{Boswellppcf1984} Such plasmas also provide a rich environment for the experimental study of drift waves and their impact on anomalous transport,\cite{Schroder2004} which is important to understand in fusion plasmas. Although there have been many publications on helicon wave physics, less attention has been devoted to the physics of low frequency oscillations in the diffusion region away from the source.\cite{Schroder2005} Light \emph{et al} \cite{Light2001, Light2002} observed a low frequency electrostatic instability only above a critical magnetic field and identified it as a mixture of drift waves and a Kelvin-Helmholtz instability. Degeling \emph{et al} \cite{Degeling1999, Degelings1999} observed relaxation oscillations in the kilohertz range that were associated with the various types of mode coupling in helicon discharges. More recently, a cylindrical linear model which treats global eigenmodes \cite{Ellis1980} has been employed to study drift waves in the magnetically confined plasma device VINETA,\cite{Schroder2005, Schroder2004} however, the model does not involve the $\mathbf{E} \times \mathbf{B}$ rotation and neglects ion parallel motion which may influence the characteristics of low frequency oscillations.\cite{Horton1984} Sutherland \emph{et al} \cite{Sutherland2005} observed low frequency ion cyclotron waves that were highly localized along the axial center of the WOMBAT plasma device. Spectral measurements revealed a four-wave interaction where energy is down-converted to the ion cyclotron mode from the helicon pump. 

Low frequency oscillations in helicon plasmas can be broadly classified into two types: the Kelvin-Helmholtz instability and drift waves. The Kelvin-Helmholtz instability is driven by a velocity shear in a mass flow or a velocity difference across the interface between two fluids. \cite{Miloshevsky2010} This can occur in helicon plasmas, because the streaming velocity of the ion fluid is much slower than that of electrons. The drift wave is a universal instability driven by a plasma pressure gradient perpendicular to the ambient field.\cite{Chenbook1984} It can arise in fully ionized, magnetically confined and low-beta plasmas,\cite{Hendel1968} and has been observed in both linear and toroidal field geometries.\cite{Okabayashi1997, Pesceli1983, Liewer1985, Klinger1992, Poli2006} In WOMBAT plasmas, the velocity shear is small and the density and temperature gradients produce large pressure gradients, suggestive of large drive of drift waves. We thus restrict attention to this class of modes. 

In this paper, a two-fluid model, which was developed originally for the vacuum arc centrifuge by Hole \emph{et al} \cite{Hole2002, Holethesis2001} to explain oscillations observed in the density and electric potential, is applied to study low frequency oscillations observed in WOMBAT. We show that the equilibrium and perturbed density profiles and the space potential profile from the model and data are consistent. The model predicts unstable modes with a similar global mode radial structure, and a frequency comparable to observed signals. The paper is organised as follows: section~\Rmnum{2} gives a brief description of the experimental setup (WOMBAT), section~\Rmnum{3} introduces the two-fluid model and section~\Rmnum{4} discusses the wave physics revealed by this model and the data. Finally, section~\Rmnum{5} discusses the possible extensions to this model and presents concluding remarks. 

\section{experimental setup}

\begin{figure}[hb]
\begin{center}$
\begin{array}{c}
\includegraphics[width=0.45\textwidth,angle=0]{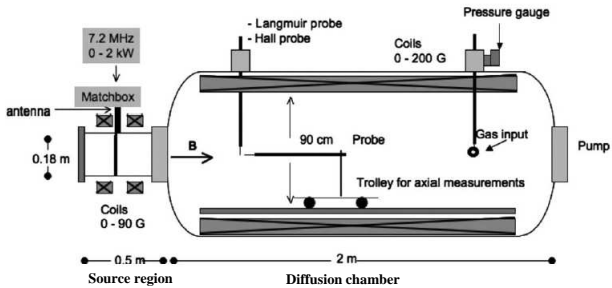}
\end{array}$
\end{center}
\caption{A schematic of the WOMBAT (Waves On Magnetised Beams And Turbulence) plasma experiment.\cite{Corr2009}}
\label{wombat}
\end{figure}

A schematic \cite{Corr2009} of the experiment is shown in Fig. ~1. The WOMBAT experiment comprises a glass source tube and a large stainless steel diffusion chamber, which are attached to each other on axis. The source tube is $50$~cm long and $18$~cm in diameter, while the diffusion chamber is $200$~cm long and $90$~cm in inner diameter. A large solenoid is employed inside the diffusion chamber to provide a steady magnetic field, which can be up to $0.02$~T and highly uniform along the axis. In this work, the large solenoid was used to produce a high density blue core. To generate the plasma, a single loop antenna ($20$~cm in diameter, $1$~cm wide and $0.3$~cm thick) was used in the source tube to couple the RF power to the argon gas. The $7.2$~MHz electric current passing through the loop creates a time varying magnetic field, which in turn induces azimuthal current in the argon gas and leads to break down and formation of the plasma.\cite{Boswellthesis1970} The base pressure of the diffusion chamber was maintained at $4 \times 10^{-6}$~Torr by a turbomolecular pump and a rotary pump. 

To measure spatial and temporal profiles in the plasma, an uncompensated, translating Langmuir probe was inserted radially into the diffusion chamber, 50 cm from the source and diffusion chamber interface. The central wire of the probe was fed through an alumina support which was in turn shielded by a 6 mm diameter grounded steel tubing covering the whole extent of the probe length up to the probe tips. The removable probe tip was made of a $0.2$~mm diameter and $8$~mm long nickel wire. For the present high density plasma discharge in the thin sheath regime (the probe sheath is much less than the probe radius), the effect brought by RF fluctuations on the floating potential and the $I(V)$ characteristics was found to be negligible. \cite{Corr2009} Radial translation of the probe was set using a computer controlled stepper motor arrangement that allowed the probe tip position to be selected with an accuracy of a few micrometers. To determine the plasma $I(V)$ characteristics, the bias voltage on the Langmuir probe was swept between $~-~50$~V and $~+~30$~V using a \emph{Labview} program. The plasma density ($n_i$), electron temperature ($T_e$), plasma potential ($V_p$) and floating potential ($V_f$) were determined from the $I(V)$ characteristics of the cylindrical probe. 

Typical Langmuir probe measurements of WOMBAT argon plasmas are shown in Table~\Rmnum{1}. Other parameters, obtained for similar helicon plasmas,\cite{Kline2003, Scime2007} and measured by laser induced fluorescence, include: the ion temperature $T_i$, the bulk rotation frequency $\omega_0$, and axial streaming velocity $V_{z0}$. The parameter $m_i$ is the argon ion mass and the charge number $Z$ is taken to match similar argon plasma conditions. \cite{Plihon2005} The remaining parameters are introduced in section~\Rmnum{3}. Finally, plasma parameters for the PCEN (Plasma CENtrifuge) device are taken from published work.\cite{Dallaqua1998, Hole2002} 

\begin{table}[ht]
\caption{\label{tab:tab1}Typical parameters of WOMBAT and PCEN (Plasma CENtrifuge).}
\begin{ruledtabular}
\begin{tabular}{lll}
parameter                                                       & WOMBAT                             & PCEN \cite{Dallaqua1998,Hole2002} \\
\hline
$n_i$ (on axis)                                                 & $1.4\times 10^{19}\ \rm{m^{-3}}$   & $5.2\times10^{19}\ \rm{m^{-3}}$\\
$T_e$                                                           & $1.5$ eV                           & $2.9$ eV\\
$T_i$                                                           & $0.1$ eV                           & $2.9$ eV\\
$m_i$                                                           & $40$ amu                           & $24.31$ amu\\
$B_z$                                                           & $0.0185$ T                         & $0.05$ T\\
$Z$                                                             & $1.0$                              & $1.5$\\
$V_{z0}$                                                        & $200\ \rm{m~s^{-1}}$               & $10^4\ \rm{m~s^{-1}}$\\
$\omega_0$                                                      & $1.42\ \rm{krad~s^{-1}}$           & $184\ \rm{krad~s^{-1}}$\\
$\omega_{ic}= \frac{B_z e Z}{m_i}$                              & $44.5\ \rm{krad~s^{-1}}$           & $295\ \rm{krad~s^{-1}}$\\
$\Omega_{i0}=\frac{\omega_0}{\omega_{ic}}$                      & $0.032$                            & $0.59$\\
$\Psi=(\frac{T_i}{T_e}+Z)\frac{k_B T_e}{m_i \omega_{ic}^2 R^2}$ & $1.72$                             & $1.6$ \\
$\delta=\frac{e Z n_{i0}}{B_z} \frac{\eta_L}{\gamma_E}$         & $0.0235$                           & $0.03$\\
$R$(characteristic radius)                                      & $6$ cm                             & $1.43$ cm
\end{tabular}
\end{ruledtabular}
\end{table}

\section{plasma model}

\subsection{Model assumptions}

The two-fluid model developed by Hole \emph{et al} \cite{Hole2002, Hole2001} is based on the following plasma assumptions:
{\setlength{\leftmargini}{12pt} 
\begin{enumerate}[topsep=0pt, partopsep=0pt]
\item Ions of different charge can be treated as a single species with average charge $Z$.
\item The plasma is quasi-neutral, giving that $n_e=Z n_i$.
\item The steady-state plasma is azimuthally symmetric and has no axial structure.
\item The effects induced by plasma fluctuations on the externally applied field can be neglected.
\item Both finite Larmor radius (FLR) and viscosity effects are negligible.
\item For the range of frequencies considered here, the electron inertia can be neglected. 
\item The ion and electron temperatures, $T_i$ and $T_e$, are uniform across the plasma column.
\item The steady-state ion density distribution has a form of $n_{0}=n_i(0)e^{-(r/R)^2}$, which is a Gaussian profile. Here, $n_i(0)$ is the on-axis ion density, and $R$ is the characteristic radius at which the density is $1/e$ of its on-axis value.
\item The steady-state velocities of ions and electrons can be written as $\mathbf{v_i}~=~(0,~\omega_i r,~\nu_{iz})$ and $\mathbf{v_e}~=~(0,~\omega_e(r) r,~\nu_{ez}(r))$, respectively, where $\omega_i$ is the ion rigid rotor rotation frequency, $\nu_{iz}$ is the ion uniform axial streaming velocity, $\omega_e(r)$ is the electron rotation frequency, and $\nu_{ez}(r)$ is the electron streaming velocity. While treated in other work \cite{Hole2001}, radial diffusion of both ions and electrons due to electron-ion collision is negligible.
\end{enumerate}
Here, length and time are normalised to $R$ and $1/\omega_{ic}$ respectively, where $\omega_{ic}=Z e B_z/m_i$ is the ion cyclotron frequency. A cylindrical coordinate system is developed, with $(x,~\theta,~\varsigma)=(r/R,~\theta,~z/R)$ and $\tau=\omega_{ic}t$, where $x$ and $\varsigma$ are the normalised radial and axial positions, respectively. 

\subsection{Two-fluid equations}

The model comprises the motion and continuity equations of ion and electron fluids, written respectively as

\begin{equation}
\frac{\partial \mathbf{u_i}}{\partial \tau}+(\mathbf{u_i} \cdot \mathbf{\nabla}) \mathbf{u_i} = -\psi(\rm{Z} \mathbf{\nabla} \chi+\lambda \mathbf{\nabla} l_i)+\mathbf{u_i} \times \mathbf{\hat{\varsigma}}+\delta \rm{n_s} \mathbf{\tilde{\xi}} \cdot (\mathbf{u_e}-\mathbf{u_i}),
\end{equation}

\begin{equation}
\psi Z(-\mathbf{\nabla} l_i+\mathbf{\nabla} \chi)-\mathbf{u_e} \times \mathbf{\hat{\varsigma}}+\delta \rm{n_s} \mathbf{\tilde{\xi}} \cdot (\mathbf{u_i}-\mathbf{u_e})=\rm{0},
\end{equation}

\begin{equation}
-\frac{\partial l_i}{\partial \tau}=\mathbf{\nabla} \cdot \mathbf{u_i}+\mathbf{u_i} \cdot \mathbf{\nabla} \rm{l_i},
\end{equation}

\begin{equation}
-\frac{\partial l_i}{\partial \tau}=\mathbf{\nabla} \cdot \mathbf{u_e}+\mathbf{u_e} \cdot \mathbf{\nabla} \rm{l_i},
\end{equation}
\\with terms defined as follows: 

\[\mathbf{u_i}=\frac{\mathbf{v_i}}{\omega_{ic} \rm{R}}=(\rm{x \varphi_i},~\rm{x \Omega_i},~\rm{u_{i\varsigma}}), \mathbf{u_e}=\frac{\mathbf{v_e}}{\omega_{ic}\rm{R}}=(\rm{x \varphi_e},~\rm{x \Omega_e},~\rm{u_{e\varsigma}}),\]

\[\lambda=\frac{T_i}{T_e}, \psi=\frac{k_B T_e}{m_i \omega_{ic}^2 R^2}, \chi=\frac{e \phi}{k_B T_e}, \mathbf{\tilde{\xi}}=diag(\xi_\bot,~\xi_\bot,~1),\]

\[l_i=ln \frac{n_i}{n_i(0)}, n_s=\frac{\nu_{ei}}{\nu_{ei}(0)}, \delta=\frac{e Z n_{i0}}{B_z} \frac{\eta_L}{\gamma_E}.\]
Here, subscript $i$ and $e$ refer to ion and electron parameters respectively, $\varphi$ is the normalised radial velocity divided by $x$, $\Omega$ is the normalised rotation frequency, $u_{\varsigma}$ indicates the normalised axial velocity, $\lambda$ is the ratio between ion and electron temperatures, $\psi$ is a convenient constant which for $\lambda=1$ becomes the square of the normalised ion thermal velocity, $\chi$ is a normalised electric potential $\phi$, $l_i$ is the logarithm of the ratio of the ion density $n_i$ to its on-axis value $n_i(0)$, and $n_s$ is the ratio of the electron-ion collision frequency $\nu_{ei}$ to its on-axis value $\nu_{ei}(0)$. Also, $\delta$ is the normalised resistivity parallel to the magnetic field, where $\eta_L$ is the electrical resistivity of a Lorentz gas and $\gamma_E$ is the ratio of the conductivity of a charge state Z to that in a Lorentz gas.\cite{Lyman1962}

\subsection{Steady-state solution}

In cylindrical geometry with a purely axial constant field ${\mathbf B}=(0,~0,~B_z)$, the steady-state solution of this model is given by
\begin{equation}
\chi_0(x)=\chi_c+[\frac{\Omega_{i0}}{2\psi Z}(1+\Omega_{i0})+\frac{\lambda}{Z}]x^2, 
\end{equation}
with
\begin{equation}
\Omega_{e0}=\Omega_{i0}(1+\Omega_{i0})+2 \psi (\lambda+Z),
\end{equation}
where $\chi_c$ is an arbitrary reference potential. The axial current in this model is unconstrained, and can arbitrarily be set to zero ($u_{i\varsigma0}~=~u_{e\varsigma0}$), consistent with WOMBAT boundary conditions. 

We have compared the steady-state solution to experimental data. While there is little spectroscopic data available to support the rigid rotor assumption, measurements of the fluctuation frequency from probe measurements suggest rigid rotation. Figure~2(a) shows the equilibrium density profile of the WOMBAT plasma for two different axial field strengths, $B_z~=~0.0185$~T and $B_z~=~0.0034$~T, respectively.\cite{Corr2007} Overlaid are best fits to the profile using a Gaussian profile ($n_{i0}~=~n_i(0)e^{-(r/R)^2}$): these show reasonable agreement over the range of the data. The stronger field produces better confinement with a smaller characteristic radius $R$.

\begin{figure}
\begin{center}$
\begin{array}{c}
\vspace{-0.7 cm}\hspace{0 cm}\includegraphics[width=0.45\textwidth,angle=0,scale=0.98]{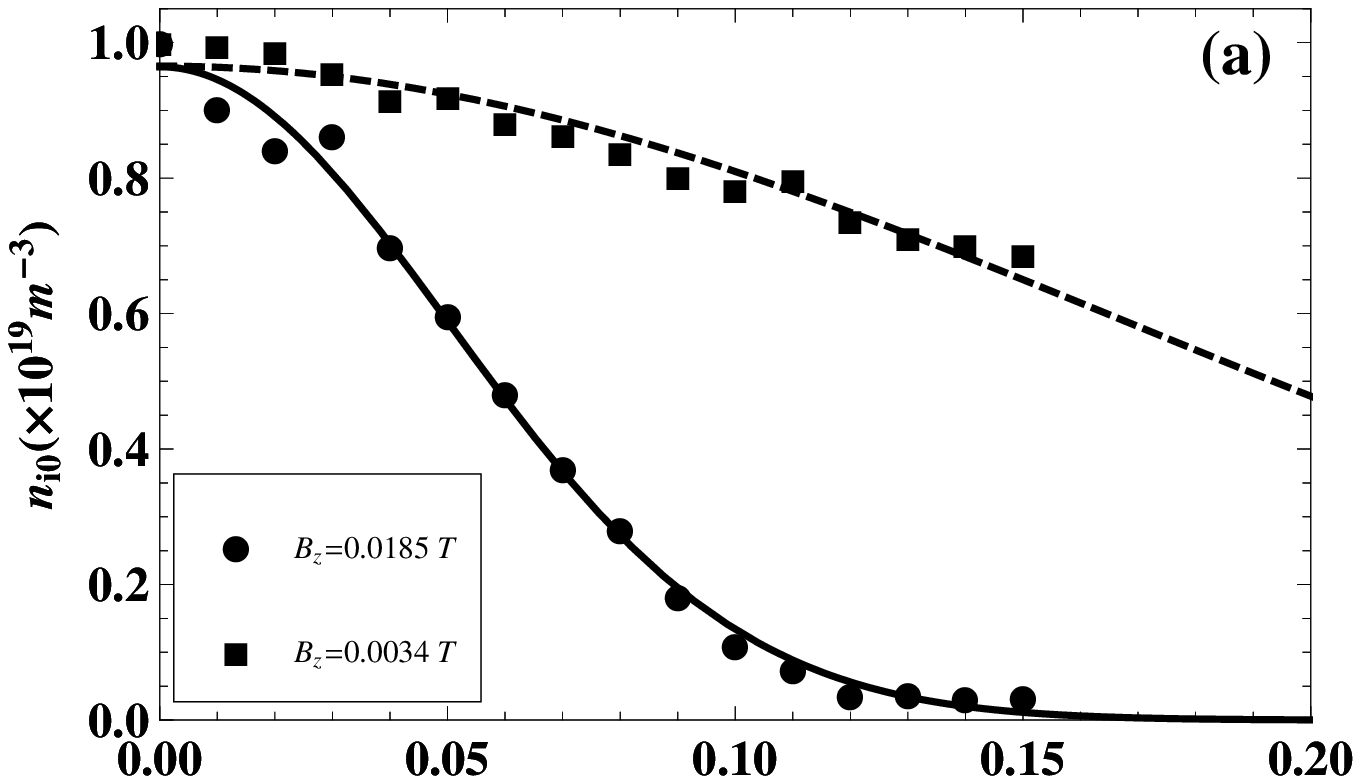}\\
\vspace{-0.7 cm}\hspace{0.63 cm}\includegraphics[width=0.45\textwidth,angle=0,scale=1.01]{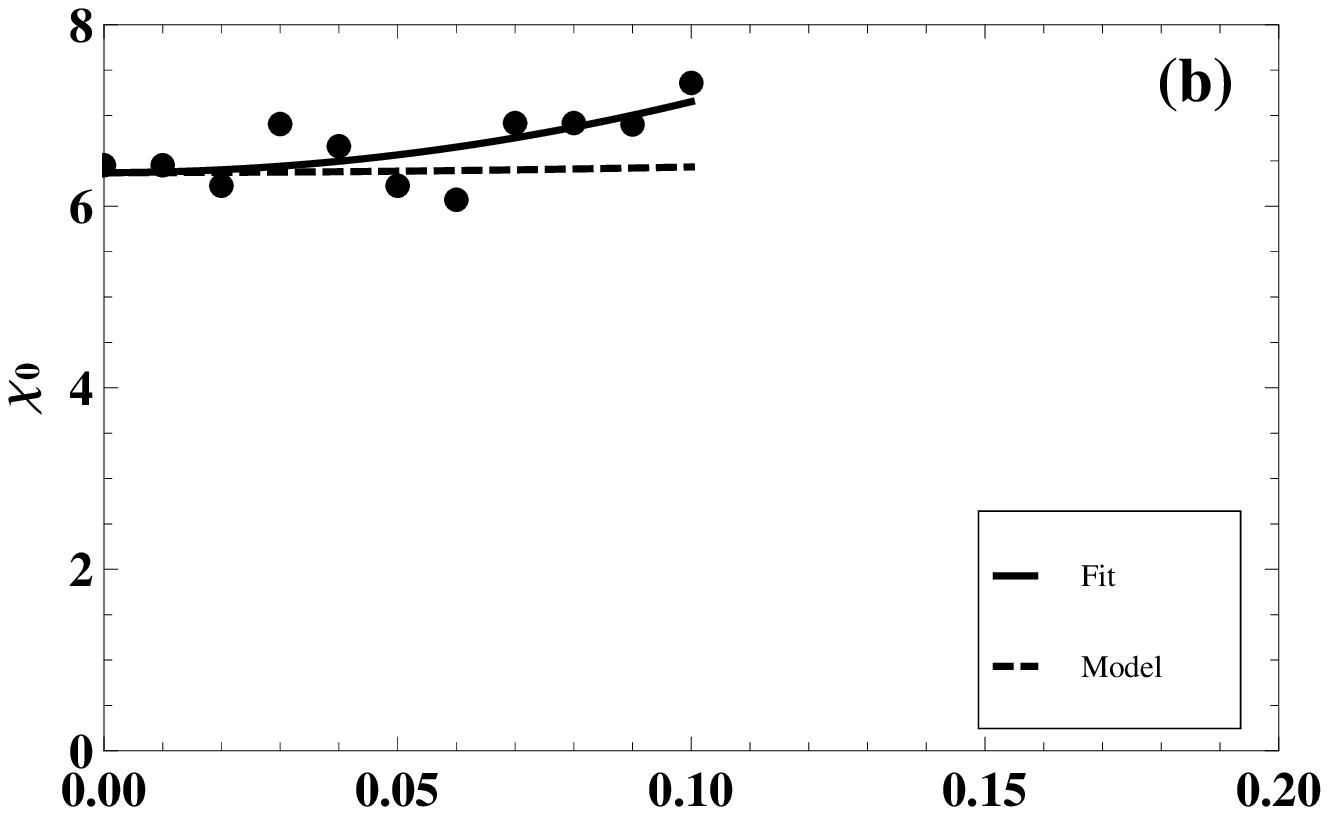}\\
\hspace{0.68 cm}\includegraphics[width=0.45\textwidth,angle=0,scale=1.025]{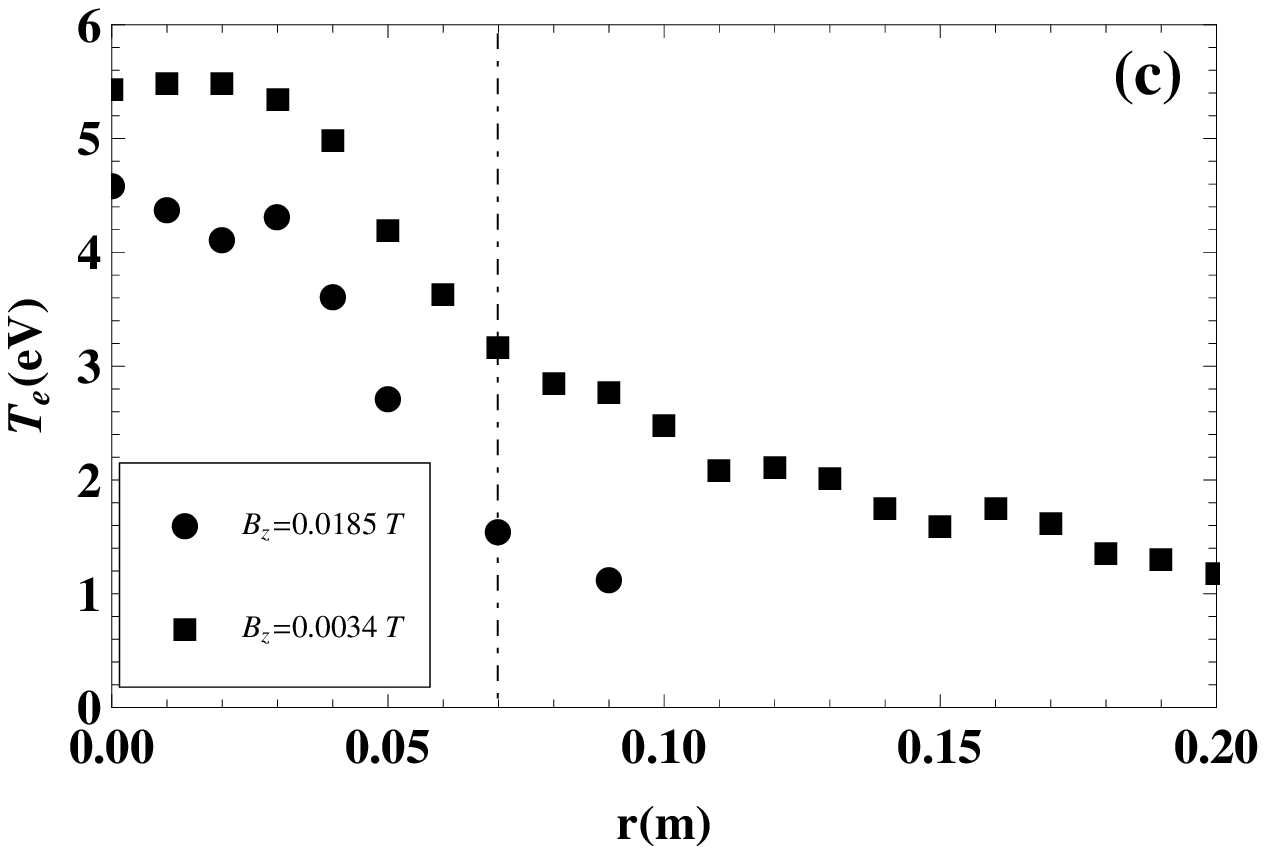}
\end{array}$
\end{center}
\caption{Plasma configuration of WOMBAT: (a) Gaussian fit of the equilibrium density profile, for $B_z~=~0.0185$~T and $B_z~=~0.0034$~T respectively;\cite{Corr2007} (b) the radial variation of the normalised space potential from the fit procedure (solid line), the model (dashed) and the data (dots),\cite{Corr2009} which was taken under the conditions of $B_z~=~0.0068$~T, $0.6$~mTorr and $1708$~W in the experiment; (c) the radial profile of the electron temperature at two field strengths: $B_z~=~0.0185$~T and $B_z~=~0.0034$~T. \cite{Corr2007} The dashed vertical line in (c) corresponds to the radial location of fluctuation measurements. }
\label{densityfit}
\end{figure}

Figure~2(b) shows the radial variation of the space potential in WOMBAT for a slightly different field strength, $B_z~=~0.0068$~T.  \cite{Corr2009} Overlaid is a best fit to the data using the parabolic potential profile in Eq.~(5), with the arbitrary reference potential ($\chi_c$) and gradient ($G~=~\frac{\Omega_{i0}}{2\psi Z}(~+~\Omega_{i0})+~\frac{\lambda}{Z}$) free parameters. Although the fit, for which $G~=~0.5$, is reasonable, the scatter in the data is large. For the same $\chi_c$, we have over plotted the model potential profile with $G~=~0.04$. Within the bulk of the plasma, out to the characteristic radius $r~=~0.08$~m, model and observed potential profiles broadly agree. 

Figure~2(c) shows the radial profile of $T_e$ in the WOMBAT plasma.\cite{Corr2007} The constant electron temperature of the model is an approximation to the radially varying experimental profile. For the model, we have chosen the $T_e$ value at the position at which the frequency was determined, $r=7$~cm. 

\subsection{Normal mode analysis}

To compute the normal modes of the system, Hole \emph{et al}\cite{Hole2002} apply a linear perturbation treatment with plasma parameters $\zeta$ taking the form

\begin{equation}
\zeta (\tau, x, \theta, \varsigma)=\zeta_0(x)+\varepsilon \zeta_1(x)e^{i(m \theta+k_\varsigma \mathbf{\varsigma}-\omega \tau)}.
\end{equation}
Here, $\varepsilon$ is the perturbation parameter, $m$ the azimuthal mode number, $k_\varsigma$ the axial wave number and $\omega$ the angular frequency. To first order in $\varepsilon$ the system of Eqs.~(1)~-~(4) reduces to

\begin{equation}
\left(
\begin{array}{clcr}
\psi (l_{i1}'(y)-X_1'(y))\\
y \varphi_{i1}'(y)\\
y \varphi_{e1}'(y)-i m \Psi X_1'(y)\\
\Psi \delta \xi_\bot e^{-y} X_1'(y)\\
0
\end{array}
\right)
=
\tilde{\mathbf{A}}
\left(
\begin{array}{clcr}
l_{i1}(y)\\
X_1(y)\\
\varphi_{i1}(y)\\
\varphi_{e1}(y)\\
u_{e\varsigma}(y)
\end{array}
\right),
\end{equation}
where $\tilde{\mathbf{A}}$ is the matrix

\begin{widetext}
\begin{equation}
\begin{array}{lll}
\tilde{\mathbf{A}} & = &
\left(
\begin{array}{ccccc}
\frac{m \Psi C}{2 \bar{\omega}y}                                                 & 0                   & \frac{i \bar{\omega}}{2}-\frac{i C^2}{2 \bar{\omega}} & \frac{i C}{2 \bar{\omega}} & 0\\
\frac{i}{2}(\bar{\omega}-\frac{\Psi}{\bar{\omega}}(\frac{m^2}{y}+k_\varsigma^2)) & 0                   & -1+y-\frac{m C}{2 \bar{\omega}}                       & \frac{m}{2 \bar{\omega}}   & 0\\
\frac{i}{2}(\bar{\omega}-m \Omega_{i0}^2-2m \Psi)                                & 0                   & 0                                                     & -1+y                       & -\frac{i k_\varsigma}{2}\\
0                                                                                & \frac{i m \Psi}{2y} & 0                                                     & -\frac{1}{2}               & 0\\
0                                                                                & -i k_\varsigma \Psi & 0                                                     & 0                          & 0\\
\end{array}
\right)\\
&+&\delta
\left(
\begin{array}{ccccc}
0                                                                                & 0 & -\frac{\xi_\bot e^{-y}}{2}                   & \frac{\xi_\bot e^{-y}}{2}                   & 0\\
0                                                                                & 0 & 0                                            & 0                                           & 0\\
0                                                                                & 0 & -\frac{i m \xi_\bot e^{-y}}{2}               & \frac{i m \xi_\bot e^{-y}}{2}               & 0\\
\xi_\bot e^{-y}(-\frac{m \Psi}{2 \bar{\omega} y}+\frac{\Omega_{i0}^2}{2}+\Psi)   & 0 & \frac{i C \xi_\bot e^{-y}}{2 \bar{\omega}}   & -\frac{i \xi_\bot e^{-y}}{2 \bar{\omega}}   & 0\\
\frac{e^{-y}k_\varsigma \Psi}{\bar{\omega}}                                      & 0 & 0                                            & 0                                           & -e^{-y}
\end{array}
\right).
\end{array}
\end{equation}
\end{widetext}

Here, $\bar{\omega}~=~\omega~-~m \Omega_{i0}~-~k_\varsigma u_{\varsigma 0}$ is the frequency in the frame of the ion fluid, $C~=~1~+~2\Omega_{i0}$, $\Psi~=~(\lambda~+~Z)\psi$, $y~=~x^2$ and we have introduced a new dependent variable $X_1(y)$, where

\begin{equation}
X_1(y)=\frac{Z}{Z+\lambda}[l_{i1}(y)-\chi_1(y)]. 
\end{equation}

For large axial wavelength modes of the resistive plasma column, for which $k_\varsigma^2~\leq~\delta$, this model can be reduced to a second order differential equation
\begin{equation}
(\frac{\bar{\omega}^2-C^2}{\bar{\omega}\Psi})L(N_c)[g_1(y)]=0,
\end{equation}
where
\[L(N_c)=y \frac{\partial^2}{\partial y^2}+(1-y)\frac{\partial}{\partial y}+(\frac{N_c}{2}-\frac{m^2}{4y}),\]
\[N_c=\frac{(\bar{\omega}^2-C^2)(m+\frac{i}{2}f(y))}{\bar{\omega}-m \Omega_{i0}^2+i \Psi f(y)}+\frac{m C}{\bar{\omega}},\]
and $f(y)~=~F^2 e^{y}$ with the normalised axial wave number $F~=~k_\varsigma/\sqrt{\delta}$. For odd $m$ modes, the boundary conditions for Eq.~(11) are $g_1(0)~=~0$ and $g_1(Y)~=~0$ with the infinite radius $Y$ representing the edge of plasma column. For even $m$, these conditions become $g_1'(0)~=~0$ and $g_1(Y)~=~0$. We seek unstable solutions to Eq.~(11), for which $\bar{\omega}^i~>~0$. The solutions $\bar{\omega}~=~\pm~C$ are stable and hence discarded.  

\section{wave physics}

The experimental results indicate that a $m~=~1$ drift wave mode dominates the density fluctuations,\cite{Shinohara2001} and that the mode appears to propagate purely azimuthally in direction of the electron diamagnetic drift. There is no clear radial component of the mode propagation. The observed phase velocity of the mode is approximately $200$~m/s in the region of maximum density perturbation. 

\subsection{Dispersion curve}

As in Hole \emph{et al},\cite{Hole2002} we have solved Eq.~(11) by a shooting method. For $m~=~1$ the boundary conditions are $g_1(0)~=~g_1(Y)~=~0$. As the differential equation is homogeneous, the gradient at the edge $g_1'(Y)$ is arbitrary: we have chosen $g_1'(Y)~=~1$. To solve for given $F$, we choose a trial $\bar{\omega}$ and march the solution from the edge to the core. The complex frequency $\bar{\omega}$ is then adjusted until the on-axis boundary condition is satisfied. We commence the procedure at $F~=~0$, for which an analytical solution for $\bar{\omega}$ is available. This is found from the solution to 

\begin{equation}
N_c=\frac{m(\bar{\omega}^2-C^2)}{\bar{\omega}-m \Omega_{i0}^2}+\frac{m C}{\bar{\omega}}=2n+|m|,
\end{equation}
where $n$ is the number of radial nodes in the plasma. 

Figure~3 shows dispersion curve for the plasma conditions of the PCEN in Table~\Rmnum{1}. We have compared the dispersion curve to Hole \emph{et al}, \cite{Hole2002} and found it to be identical, thus validating our numerics. For PCEN plasmas, the peak of normalised growth rate $\bar{\omega}^i~=~0.39$ lies at $F~=~0.3$, with normalised frequency $\bar{\omega}^r~=~0.007$: the wave is thus near stationary in the frame of the ion fluid. The mode crossings located at $F~=~0.55$ are associated with the centrifugal instability. Also shown is the dispersion curve for WOMBAT plasma conditions of Table~\Rmnum{1}. For WOMBAT plasmas, the peak growth rate of $\bar{\omega}^i~=~0.14$ occurs at $F~=~0.37$, for which $\bar{\omega}^r~=~0.44$. There is no mode crossing in the dispersion curve for WOMBAT plasmas. Also, $\bar{\omega}^r~>~0$, and so in the laboratory frame $\omega^r~=~\bar{\omega}^r~+~m \Omega_{i0}~+~k_\varsigma u_{\varsigma 0}$, approximately $\omega^r~\approx~\bar{\omega}^r~+~\Omega_{i0}$, such that the frequency of unstable waves is always larger than the sum of the plasma rotation frequency and axial velocity, and the wave propagates in the $+~\theta$ direction (electron diamagnetic drift direction). We attribute the difference between dispersion curves to the very low ion temperature $T_i$ and slow normalised rotation frequency $\Omega_{i0}$ in WOMBAT plasmas, compared with PCEN plasmas. 

\begin{figure}
\begin{center}$
\begin{array}{c}
\vspace{-0.5 cm}\includegraphics[width=0.45\textwidth,angle=0]{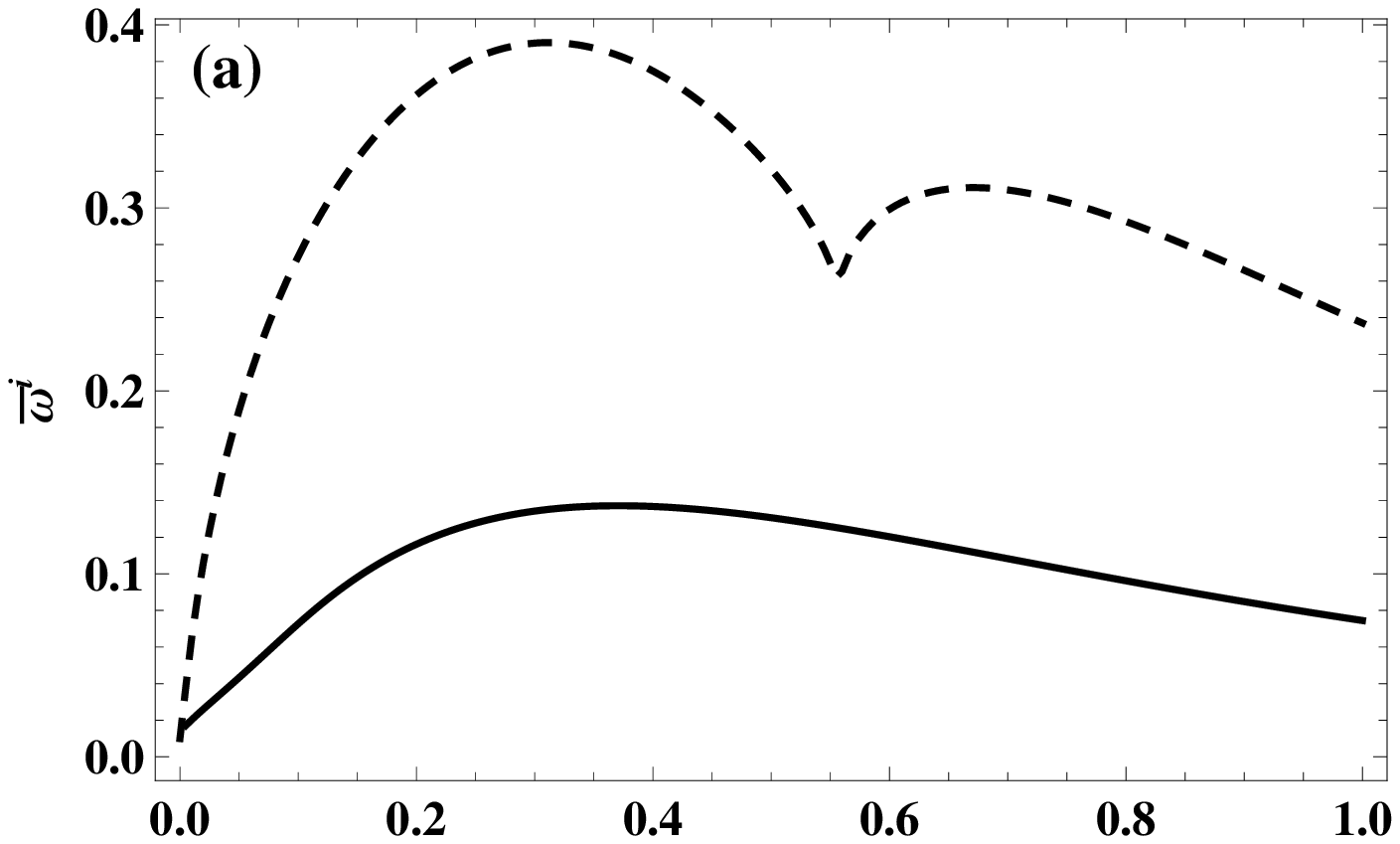}\\
\hspace{-0.65 cm}\includegraphics[width=0.45\textwidth,angle=0]{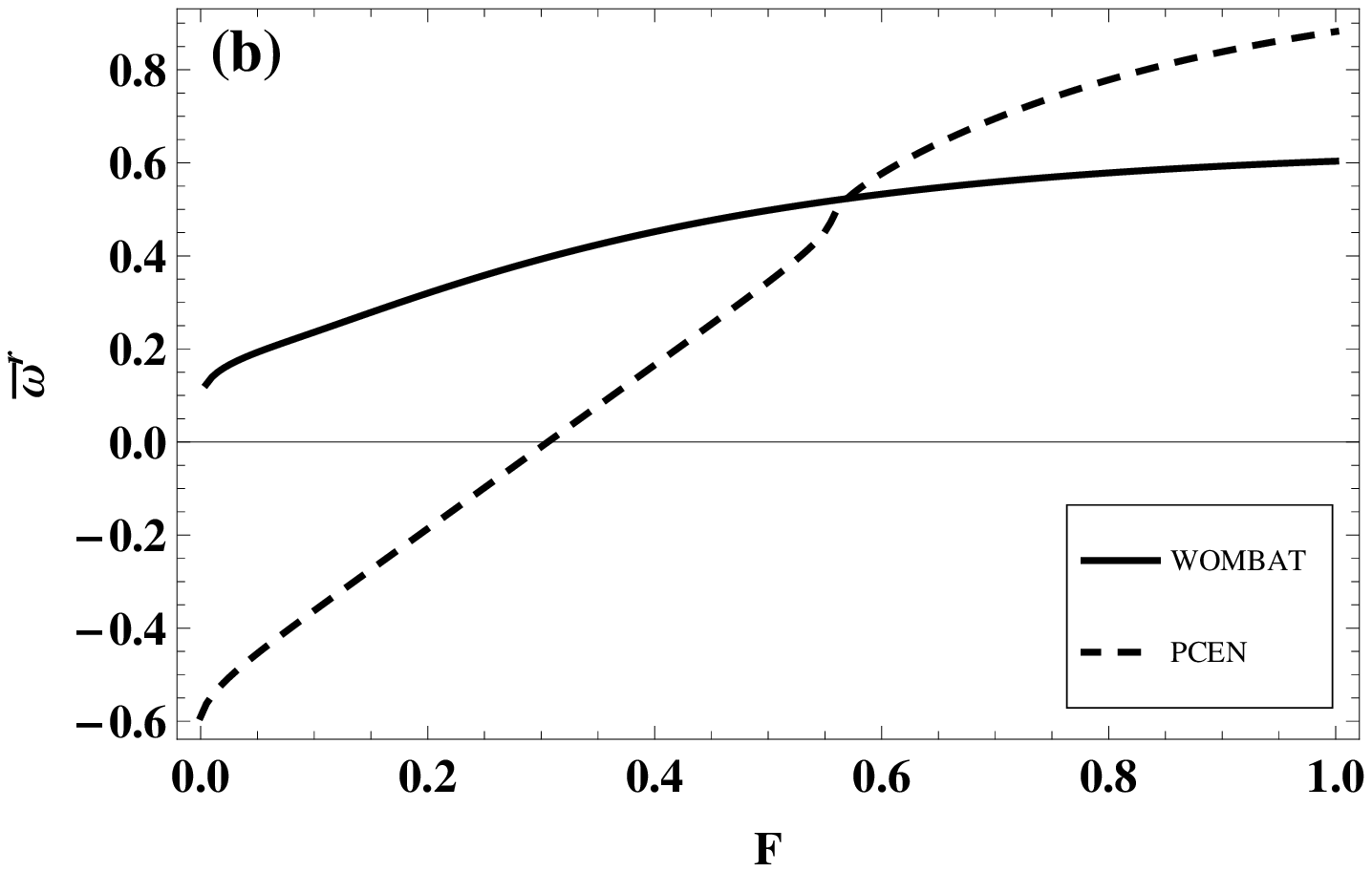}
\end{array}$
\end{center}\caption{Dispersion curves of PCEN (dashed line) and WOMBAT (solid line), generated by the model based on the conditions shown in Table~\Rmnum{1}: (a) normalised growth rate $\bar{\omega}^i$; (b) normalised frequency $\bar{\omega}^r$. (vs normalised axial wavenumber $F~=~k_\varsigma/\sqrt{\delta}$).}
\label{dispersion}
\end{figure}

\subsection{Dispersion curve sensitivity with plasma parameters}

\begin{figure*}[ht]
\begin{center}$
\begin{array}{cc}
\hspace{-0.58 cm} \vspace{-0.2 cm}\includegraphics[width=0.45\textwidth,angle=0, scale=0.927]{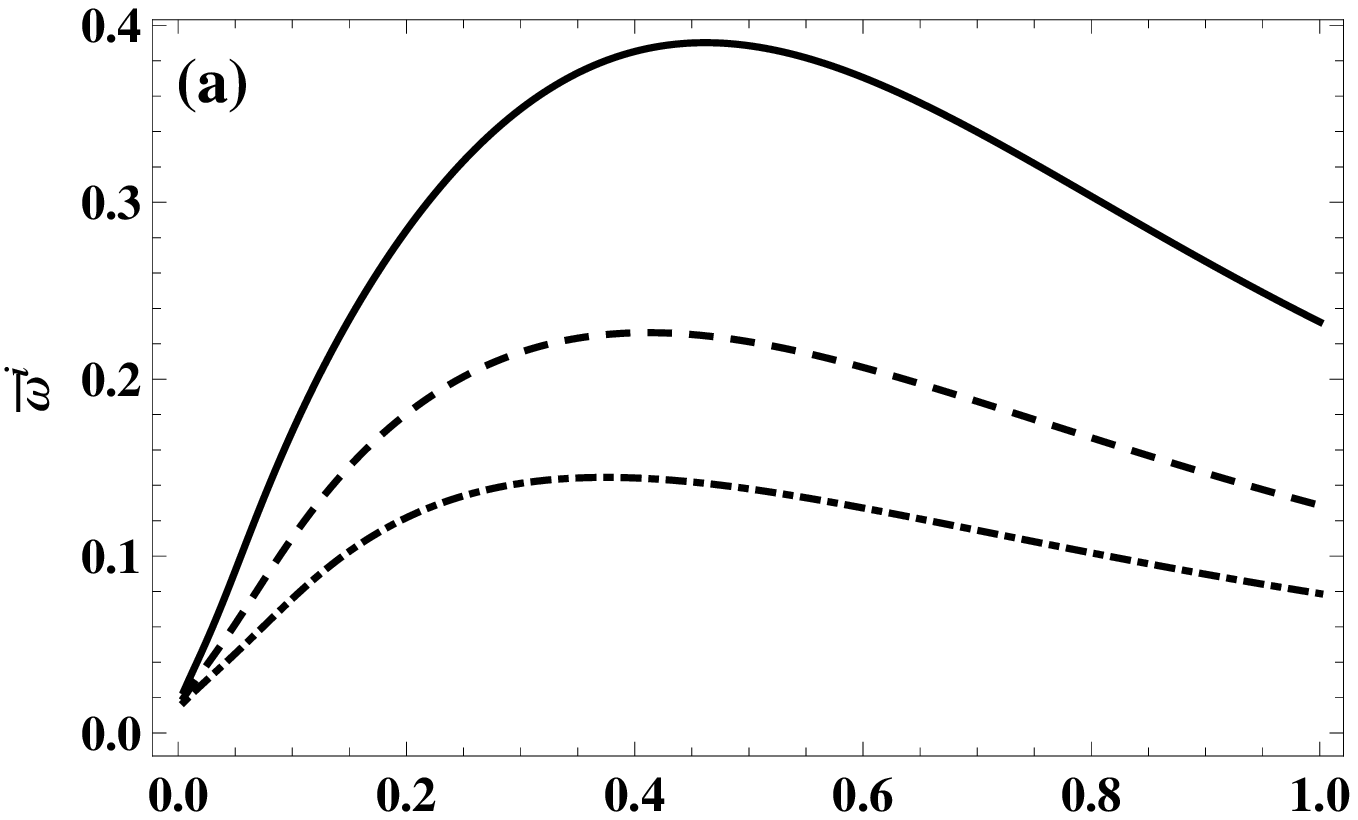}             & \vspace{-0.2 cm}\includegraphics[width=0.45\textwidth,angle=0, scale=0.965]{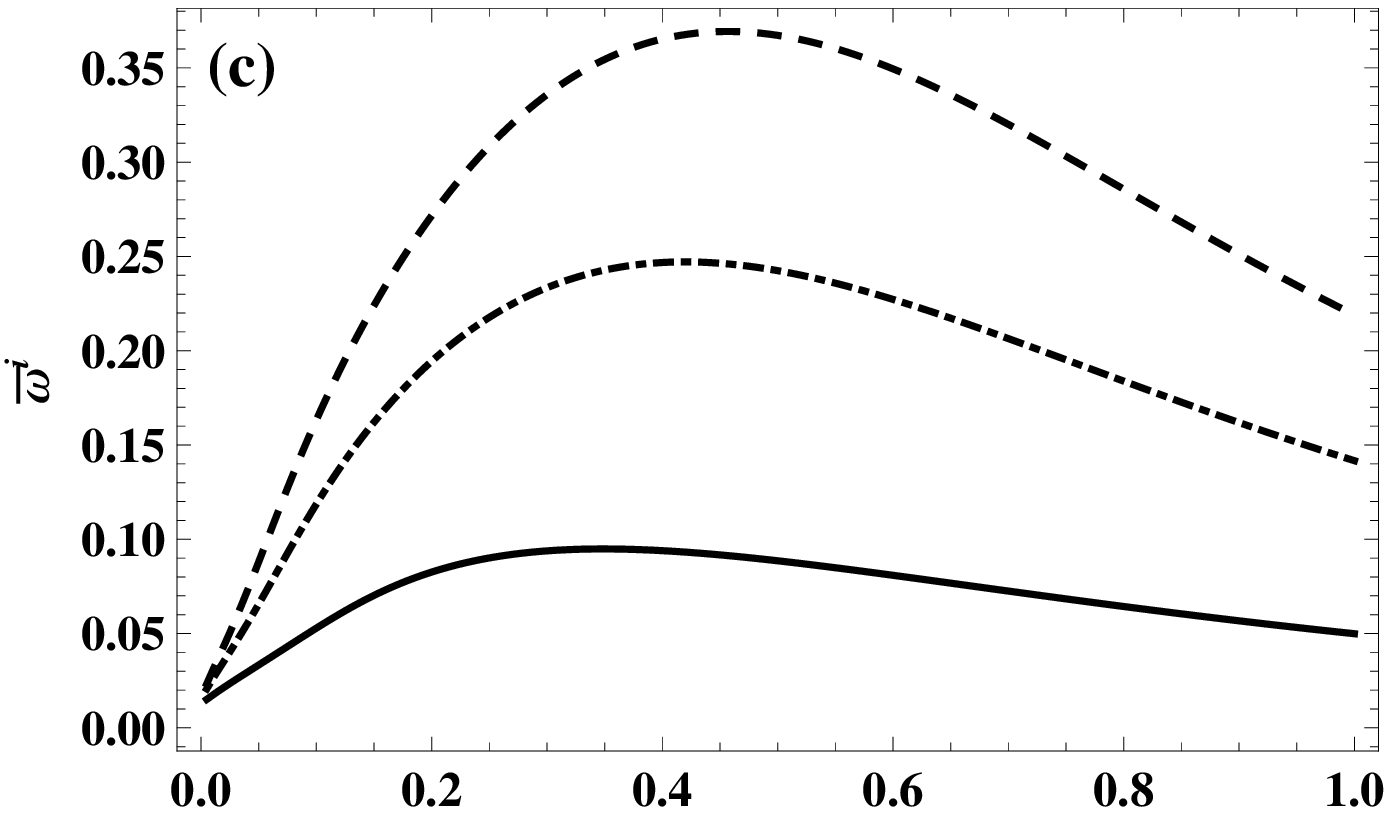}\\
\includegraphics[width=0.45\textwidth,angle=0]{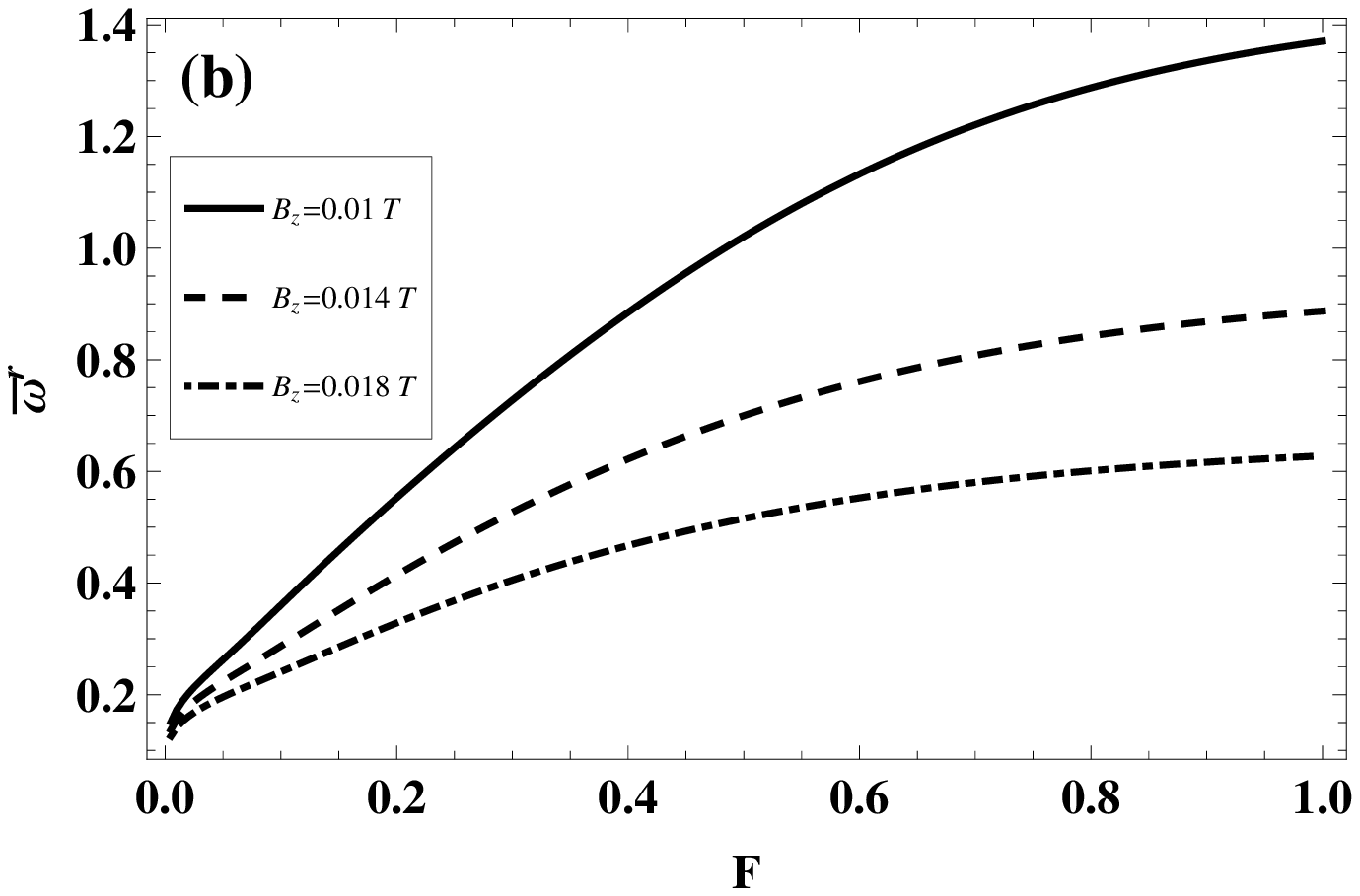} & \hspace{0.55 cm}\includegraphics[width=0.45\textwidth,angle=0]{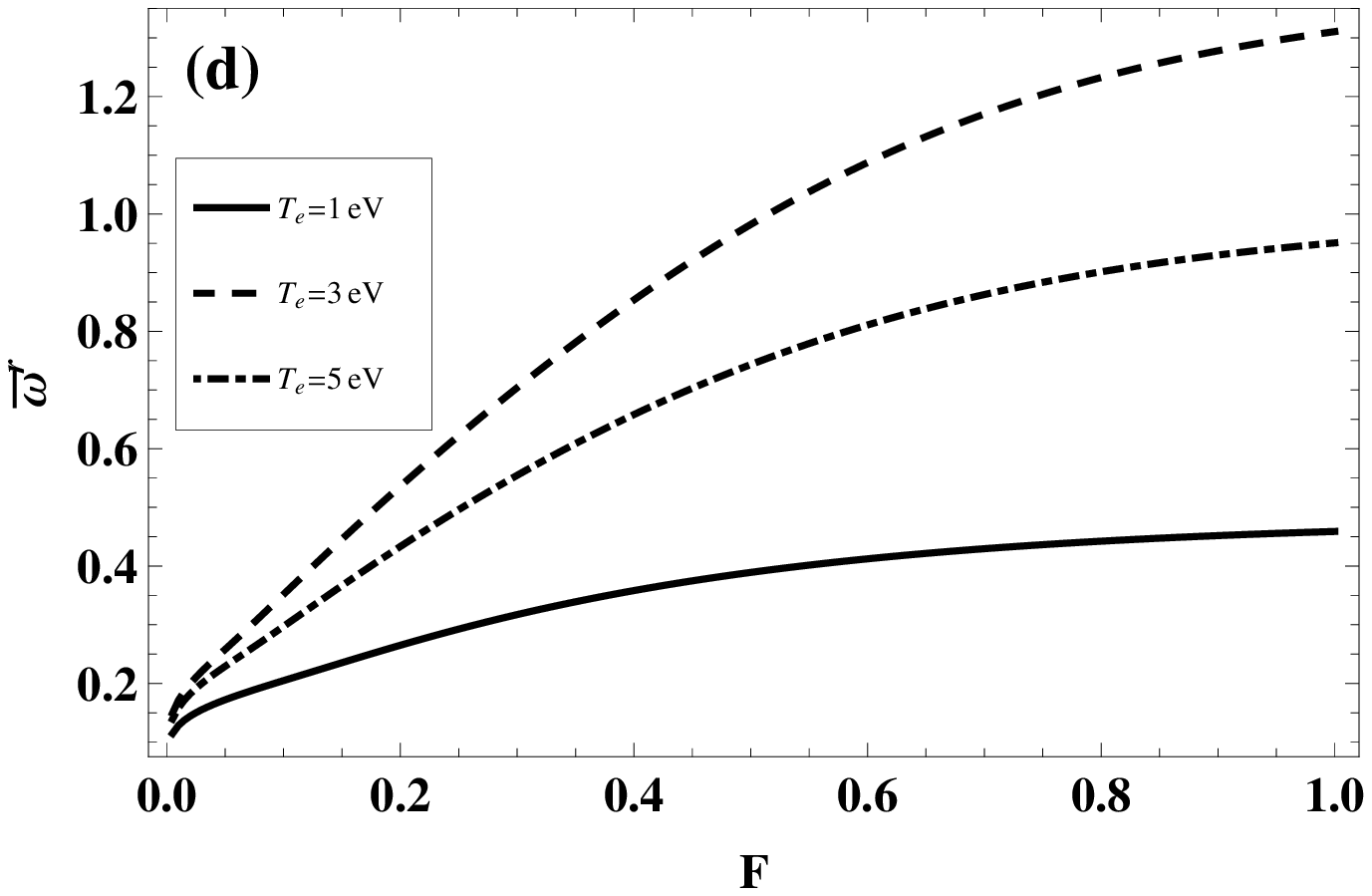}
\end{array}$
\end{center}
\caption{Results of dispersion curve sensitivity with $B_z$ and $T_e$: (a), (b) scanning results of $B_z$; (c), (d) scanning results of $T_e$.}
\label{parameter}
\end{figure*}

We have examined the sensitivity of the WOMBAT plasma dispersion curve with $\Omega_{i0}$, $B_z$ and $T_e$. For $\Omega_{i0}~<~0.05$ the dispersion curve changes by less than $1~\%$. Figure~4 shows the change in dispersion curve for different $B_z$ and $T_e$. Figure~4(a), (b) shows that $\bar{\omega}^i$ and $\bar{\omega}^r$ decrease with increasing $B_z$: the physical growth rate $\omega^i~=~\bar{\omega}^i \omega_{ic}$ also obeys this trend while $\omega^r~=~\bar{\omega}^r \omega_{ic}$ exhibits the same trend in the range $F~>~0.5$. Also, the peak growth rate shifts to lower $k_\varsigma$, and so for constant $R$ the axial wavelength of the most unstable mode is increased. In contrast, Fig.~4(c), (d) shows that $\bar{\omega}^i$ and $\bar{\omega}^r$ increase with increasing $T_e$, and the peak growth rate shifts to larger $k_\varsigma$. The dependence of growth rate with pressure gradient, and frequency with inverse field strength is consistent with a resistive drift wave.\cite{Chenbook1984} 

\subsection{Wave oscillation frequency with $B_z$ and $T_e$}

In this section we compute the variation in wave frequency at the maximum growth rate with field strength, for two choices of electron temperature: a constant $T_e~=~2.13$~eV, independent of $B_z$; and a $B_z$ dependent $T_e$. As $T_e$ data is only available at two field strengths (Fig.~2(c)):  $B_z~=~0.0185$~T, $T_e~=~3.2$~eV; and $B_z~=~0.0034$~T, $T_e~=~1.5$~eV, we have used linear interpolation to compute $T_e$ at other field strengths. Fluctuation data is available at five field strengths: $B_z~=~0.0129$~T, $B_z~=~0.0146$~T, $B_z~=~0.0162$~T, $B_z~=~0.0179$~T and $B_z~=~0.0195$~T. We have solved the dispersion curve for each case, and selected $\bar{\omega}^r$ corresponding to the maximum $\bar{\omega}^i$ to calculate the frequency in the laboratory frame. 

Figure~5 shows the measured and predicted oscillation frequencies. For both curves, the predicted and measured frequencies differ by a factor of $3.5$. There are various possible reasons for this gap. The most likely is that the model assumes flat $T_e$ and $T_i$ profiles, whereas WOMBAT plasmas exhibit a nonuniform $T_e$ profile. It may also be the case that the $T_i$ profile is nonuniform. Second, there is little data to substantiate the rigid rotor assumption of WOMBAT, \cite{Scime2007} and there may also be slip between the rotation of WOMBAT plasmas and the oscillation frequency. Third, we have neglected the fluctuations in the magnetic field, which may affect the dispersion curve. Finally, the calculation of $T_e$ for each $B_z$ is a linear interpretation between the two known data, which may bring errors and imprecisions into the prediction. 

\begin{figure}
\begin{center}$
\begin{array}{c}
\includegraphics[width=0.45\textwidth,angle=0, scale=1.05]{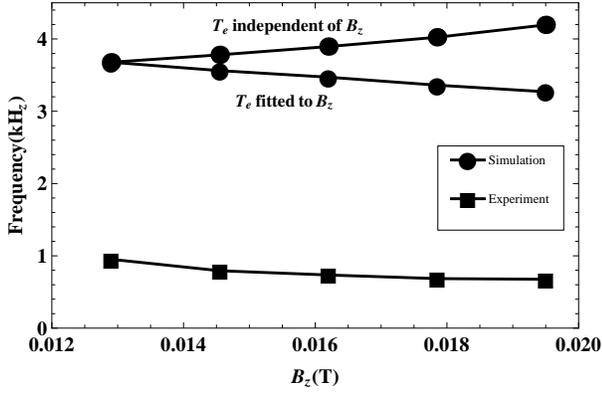}
\end{array}$
\end{center}
\caption{Wave oscillation frequency with $B_z$ and $T_e$}
\label{frequency}
\end{figure}

Figure~5 also reveals that two trends predicted by using a constant $T_e~=~2.13$~eV and a $B_z$ dependent $T_e$ are divergent. Predictions for the $B_z$ dependent $T_e$ profile exhibit the same trend as the data, suggesting the model may have correctly captured the $T_e$ dependence with $B_z$. 

Finally, to ensure that this model is in principle capable to reproduce the data, we adjusted $T_e$ to fit the data points. We found that a value of $T_e~=~0.5$~eV at low field dropping to $T_e~=~0.3$~eV at high field was able to reproduce the observed frequency. 

\subsection{Perturbed density profile}

\begin{figure}
\begin{center}$
\begin{array}{c}
\includegraphics[width=0.48\textwidth,angle=0, scale=1.05]{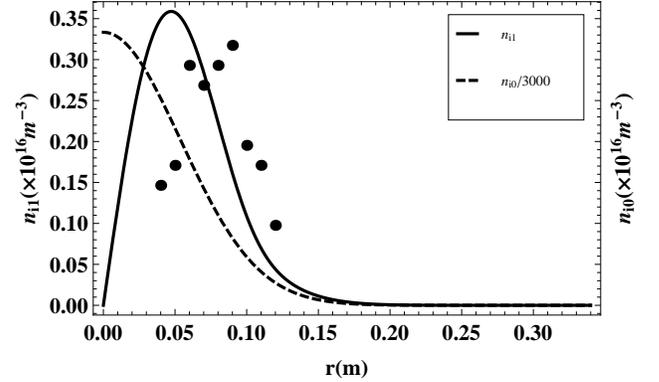}
\end{array}$
\end{center}
\caption{Comparison between the measured (dots) and predicted (solid) radial profiles of the perturbed density $n_{i1}$, together with the equilibrium density $n_{i0}$ (dashed).}
\label{perturbation}
\end{figure}

Figure~6 shows the measured and predicted radial profiles of the perturbed density $n_{i1}$, together with the equilibrium density $n_{i0}$. The data used here were taken under the conditions of $B_z~=~0.01$~T, $R~=~7.6$~cm and $T_e~=~2.46$~eV. The linear perturbation treatment gives no solution for the absolute magnitude of $n_{i1}$, and so we have fitted for the amplitude. Inspection of Fig.~6 suggests $n_{i1}$ has a single peak, and therefore consistent with $n=0$ of the perturbed mode. Also, the peak of the eigenfunction occurs in the region where the equilibrium density gradient is large, suggesting that the instability has resistive drift-type characteristics. \cite{Hendel1968} The exact radial position of the peak in $n_{i1}$ however does not agree. This could be due to uncertainty in the core position of the plasma, or the assumption of constant $T_e$ in the model.

Finally, to complement our visualisation of the mode structure, we have computed the vector field of the linear perturbed mass flow by $m_i (n_{i1} \mathbf{u_{i0}}~+~n_{i0} \mathbf{u_{i1}})$. The perturbed velocity components ${\mathbf u_{i1}}~=~(\rm{x \varphi_{i1}},~\rm{x \Omega_{i1}},~\rm{u_{i\varsigma1}})$,${\mathbf u_{e1}}~=~(\rm{x \varphi_{e1}},~\rm{x \Omega_{e1}},~\rm{u_{e\varsigma1}})$ and perturbed density $n_{i1}$ can be computed from the solution of $g_1(y)$ and following equations:

\begin{equation}
l_{i1}(y)=\frac{-g_1(y)}{(1+\frac{i}{\Psi}(\frac{m \Omega_{i0}^2+2m \Psi-\bar{\omega}}{f(y)-2 i m}))},
\end{equation}

\begin{equation}
\chi_1(y)=-\frac{\lambda}{Z}l_{i1}(y)-(1+\frac{\lambda}{Z})g_1(y),
\end{equation}

\begin{equation}
\setlength{\extrarowheight}{0.3cm}
\begin{array}{rcl}
\varphi_{i1}(y) & = & \frac{2\bar{\omega}}{i(\bar{\omega}^2-C^2)-\bar{\omega} \delta \xi_\bot e^{-y}}[\Psi(l_{i1}'(y)-X_1'(y))\\
&&-\frac{m \Psi C}{2\bar{\omega}y}l_{i1}(y)-(\frac{i C}{2\bar{\omega}}+\frac{\delta \xi_\bot e^{-y}}{2})\varphi_{e1}(y)],
\end{array}
\end{equation}

\begin{equation}
\setlength{\extrarowheight}{0.3cm}
\begin{array}{rcl}
\varphi_{e1}(y) & = &\frac{2\bar{\omega}}{\bar{\omega}+i \delta \xi_\bot e^{-y}}[\frac{i m \Psi}{2y}X_1(y)-\Psi \delta \xi_\bot e^{-y} X_1'(y)\\
&&+\delta \xi_\bot e^{-y}(-\frac{m \Psi}{2 \bar{\omega}y}+\frac{\Omega_{i0}^2}{2}+\Psi)l_{i1}(y)\\
&&+\frac{i C \delta \xi_\bot e^{-y}}{2\bar{\omega}}\varphi_{i1}(y)],
\end{array}
\end{equation}

\begin{equation}
\Omega_{i1}(y)=\frac{1}{\bar{\omega}}[m \Psi \frac{l_{i1}(y)}{y}+i(\varphi_{e1}(y)-C\varphi_{i1}(y))], 
\end{equation}

\begin{equation}
u_{iz1}(y)=\frac{\sqrt{\delta} F \Psi}{\bar{\omega}}l_{i1}(y).
\end{equation}

Figure~7 shows the flow vector field at time $t~=~0$. As time advances, the mass vector field rotates in the clockwise direction, which is the direction of electron diamagnetic drift with $B_z$ into the page. Here, the coordinate $x$ and $y$ label the cross-section of the plasma, with $x~=~r\cos\theta$ and $y~=~r\sin\theta$. The $\theta$ vector points are clockwise with $z$ into the page. The mass flow is zero on axis ($x~=~0,~y~=~0$), consistent with the boundary conditions for density. 

\begin{figure}[ht]
\begin{center}$
\begin{array}{c}
\includegraphics[width=0.45\textwidth,angle=0]{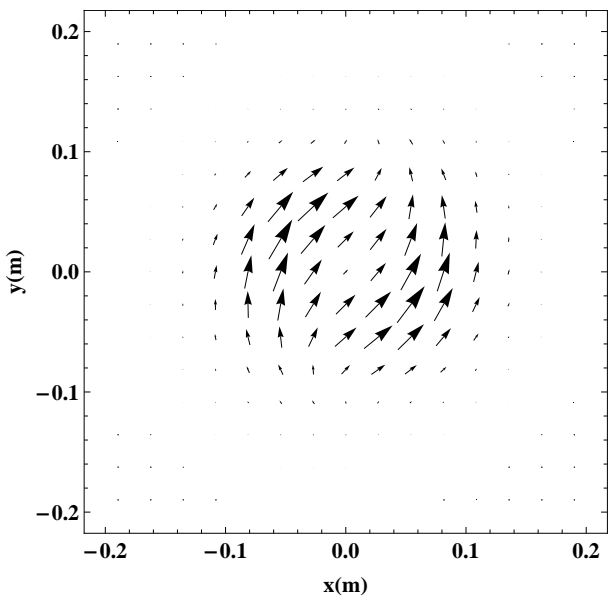}
\end{array}$
\end{center}
\caption{Vector field plot of the real part of the perturbed mass flow for mode $(m=1, n=0)$ in a cross-section of the WOMBAT plasma.}
\label{vector}
\end{figure}

\section{Conclusions}

In this paper, we employed a two-fluid model, which was developed originally for describing the wave oscillations observed in PCEN, \cite{Hole2002, Hole2001, Holethesis2001} to study the low frequency oscillations in WOMBAT. To ensure that this model is consistent with WOMBAT, the measured and predicted plasma configurations were compared, including the equilibrium density profile and the space potential profile. These show that although the density and space potential profiles agree, the temperature profile is not flat, as assumed by the model. Next, dispersion curves were generated for WOMBAT plasmas. Compared to the more rapidly rotating centrifuge plasma, the drift wave instability, unstable at larger wavelengths, has a normalised frequency much larger than the normalised frequency for the centrifuge. The difference between dispersion curves is principally due to the low rotation speed of WOMBAT plasmas compared to those of the centrifuge. 

Our study of the wave oscillation frequency with $B_z$ and $T_e$ reveals that the measured fluctuation frequency is a factor of $3.5$ lower than the predicted frequency, and the predicted trend of oscillation frequency with $B_z$ for inferred $T_e$ matches the data. The discrepancy between the measured and predicted fluctuation frequencies may be attributable to the limitations of assumed temperature uniformity across plasma column. Weaker model limitations may include the possibility the plasma does not rotate rigidly and our neglect of effects induced by plasma fluctuations on the externally applied field. Data limitations include the simultaneous measurements of plasma rotation profile and probe frequency, and the measured dependency of $T_e$ with $B_z$. 

Finally, we find that the measured and predicted perturbed density profiles have a single peak in the radial direction, indicating the perturbed mode has $n~=~0$, and the peak position is in the region of maximum equilibrium density gradient. This qualitative agreement consolidates earlier claims that the mode is a drift mode, driven by the density gradient of the plasma. To our knowledge, this is the first detailed physics model of flowing plasmas in the diffusion region away from the RF source.


\begin{thebibliography}{10}

\bibitem{Rostoker1961}
N.~Rostoker and A.~C. Kolb.
\newblock Fission of a hot plasma.
\newblock {\em Physical Review}, 124(4):965--969, 1961.

\bibitem{Freidberg1978}
J.~P. Freidberg and L.~D. Pearlstein.
\newblock Rotational instabilities in a theta-pinch.
\newblock {\em Physics of Fluids}, 21(7):1207--1217, 1978.

\bibitem{Perkins1963}
W.~A. Perkins and R.~F. Post.
\newblock Observation of plasma instability with rotational effects in a mirror
  machine.
\newblock {\em Physics of Fluids}, 6(11):1537--1558, 1963.

\bibitem{Kuo1964}
L.~G. Kuo, E.~G. Murphy, M.~Petravic, and D.~R. Sweetman.
\newblock Experimental and theoretical studies of instabilities in a
  high-energy neutral injection mirror machine.
\newblock {\em Physics of Fluids}, 7(7):988--1000, 1964.

\bibitem{Hooper1983}
E.~B. Hooper, G.~A. Hallock, and J.~H. Foote.
\newblock Low-frequency oscillations in the central cell of the tmx tandem
  mirror experiment.
\newblock {\em Physics of Fluids}, 26(1):314--322, 1983.

\bibitem{Motley1963}
R.~W. Motley and N.~Dangelo.
\newblock Excitation of electrostatic plasma oscillations near the ion
  cyclotron frequency.
\newblock {\em Physics of Fluids}, 6(2):296--299, 1963.

\bibitem{Krishnan1981}
M.~Krishnan, M.~Geva, and J.~L. Hirshfield.
\newblock Plasma centrifuge.
\newblock {\em Physical Review Letters}, 46(1):36--38, 1981.

\bibitem{Boswell1983}
R.~W. Boswell and P.~J. Kellogg.
\newblock Characteristics of 2 types of beam-plasma-discharge in a laboratory
  experiment.
\newblock {\em Geophysical Research Letters}, 10(7):565--568, 1983.

\bibitem{Boswell1984}
R.~W. Boswell, S.~M. Hamberger, P.~J. Kellogg, I.~Morey, and R.~K. Porteous.
\newblock Direct observation of rapid impulsive electron heating during a
  beam-plasma interaction.
\newblock {\em Physics Letters A}, 101(9):501--504, 1984.

\bibitem{Degeling1999}
A.~W. Degeling, T.~E. Sheridan, and R.~W. Boswell.
\newblock Model for relaxation oscillations in a helicon discharge.
\newblock {\em Physics of Plasmas}, 6(5):1641--1648, 1999.

\bibitem{Greiner1999}
F.~Greiner, O.~Grulke, H.~Thomsen, C.~Lechte, T.~Klinger, and A.~Piel.
\newblock {\em Observation of coherent structures in the turbulent equilibrium
  of a toroidal helicon discharge}.
\newblock International Conference on Phenomena in Ionized Gas, Vol I,
  Proceedings. 1999.

\bibitem{Sun2005}
X.~Sun, C.~Biloiu, and E.~Scime.
\newblock Observation of resistive drift alfven waves in a helicon plasma.
\newblock {\em Physics of Plasmas}, 12(10):--, 2005.

\bibitem{Boswell1987}
R.~W. Boswell and R.~K. Porteous.
\newblock Large volume, high-density rf inductively coupled plasma.
\newblock {\em Applied Physics Letters}, 50(17):1130--1132, 1987.

\bibitem{Ellingboe1996}
A.~R. Ellingboe and R.~W. Boswell.
\newblock Capacitive, inductive and helicon-wave modes of operation of a
  helicon plasma source.
\newblock {\em Physics of Plasmas}, 3(7):2797--2804, 1996.

\bibitem{Corr2007}
C.~S. Corr and R.~W. Boswell.
\newblock High-beta plasma effects in a low-pressure helicon plasma.
\newblock {\em Physics of Plasmas}, 14(12), 2007.

\bibitem{Boswellppcf1984}
R.~W. Boswell.
\newblock Very efficient plasma generation by whistler waves near the lower
  hybrid frequency.
\newblock {\em Plasma Physics and Controlled Fusion}, 26(10):1147--1162, 1984.

\bibitem{Schroder2004}
C.~Schroder, O.~Grulke, T.~Klinger, and V.~Naulin.
\newblock Spatial mode structures of electrostatic drift waves in a collisional
  cylindrical helicon plasma.
\newblock {\em Physics of Plasmas}, 11(9):4249--4253, 2004.

\bibitem{Schroder2005}
C.~Schroder, O.~Grulke, T.~Klinger, and V.~Naulin.
\newblock Drift waves in a high-density cylindrical helicon discharge.
\newblock {\em Physics of Plasmas}, 12(4), 2005.

\bibitem{Light2001}
M.~Light, F.~F. Chen, and P.~L. Colestock.
\newblock Low frequency electrostatic instability in a helicon plasma.
\newblock {\em Physics of Plasmas}, 8(10):4675--4689, 2001.

\bibitem{Light2002}
M.~Light, F.~F. Chen, and P.~L. Colestock.
\newblock Quiescent and unstable regimes of a helicon plasma.
\newblock {\em Plasma Sources Science and Technology}, 11(3):273--278, 2002.

\bibitem{Degelings1999}
A.~W. Degeling, T.~E. Sheridan, and R.~W. Boswell.
\newblock Intense on-axis plasma production and associated relaxation
  oscillations in a large volume helicon source.
\newblock {\em Physics of Plasmas}, 6(9):3664--3673, 1999.

\bibitem{Ellis1980}
R.~F. Ellis, E.~Mardenmarshall, and R.~Majeski.
\newblock Collisional drift instability of a weakly ionized argon plasma.
\newblock {\em Plasma Physics and Controlled Fusion}, 22(2):113--131, 1980.

\bibitem{Horton1984}
W.~Horton and J.~Liu.
\newblock Drift waves in rotating plasmas.
\newblock {\em Physics of Fluids}, 27(8):2067--2075, 1984.

\bibitem{Sutherland2005}
O.~Sutherland, M.~Giles, and R.~Boswell.
\newblock Ion cyclotron production by a four-wave interaction with a helicon
  pump.
\newblock {\em Physical Review Letters}, 94(20):--, 2005.

\bibitem{Miloshevsky2010}
G.~V. Miloshevsky and A.~Hassanein.
\newblock Modelling of kelvin-helmholtz instability and splashing of melt
  layers from plasma-facing components in tokamaks under plasma impact.
\newblock {\em Nuclear Fusion}, 50(11):1--12, 2010.

\bibitem{Chenbook1984}
Francis~F. Chen.
\newblock {\em Introduction to plasma physics and controlled fusion}, volume
  Volume 1: Plasma Physics.
\newblock Plenum Press, New York, 1984.

\bibitem{Hendel1968}
H.~W. Hendel, T.~K. Chu, and P.~A. Politzer.
\newblock Collisional drift waves-identification stabilization and enhanced
  plasma transport.
\newblock {\em Physics of Fluids}, 11(11):2426--2439, 1968.

\bibitem{Okabayashi1997}
M.~Okabayashi and V.~Arunasalam.
\newblock Study of drift-wave turbulence by microwave-scattering in a toroidal
  plasma.
\newblock {\em Nuclear Fusion}, 17(3):497--513, 1977.

\bibitem{Pesceli1983}
H.~L. Pecseli, T.~Mikkelsen, and S.~E. Larsen.
\newblock Drift wave turbulence in low-beta plasmas.
\newblock {\em Plasma Physics and Controlled Fusion}, 25(11):1173--1197, 1983.

\bibitem{Liewer1985}
P.~C. Liewer.
\newblock Measurements of microturbulence in tokamaks and comparisons with
  theories of turbulence and anomalous transport.
\newblock {\em Nuclear Fusion}, 25(5):543--621, 1985.

\bibitem{Klinger1992}
T.~Klinger and A.~Piel.
\newblock Investigations of attractors arising from the interaction of drift
  waves and potential relaxation instabilities.
\newblock {\em Physics of Fluids B-Plasma Physics}, 4(12):3990--3995, 1992.

\bibitem{Poli2006}
F.~M. Poli, S.~Brunner, A.~Diallo, A.~Fasoli, I.~Furno, B.~Labit, S.~H. Muller,
  G.~Plyushchev, and M.~Podesta.
\newblock Experimental characterization of drift-interchange instabilities in a
  simple toroidal plasma.
\newblock {\em Physics of Plasmas}, 13(10):--, 2006.

\bibitem{Hole2002}
M.~J. Hole, R.~S. Dallaqua, S.~W. Simpson, and E.~Del~Bosco.
\newblock Plasma instability of a vacuum arc centrifuge.
\newblock {\em Physical Review E}, 65(4):--, 2002.

\bibitem{Holethesis2001}
M.~J. Hole.
\newblock {\em Analysis of the Rotating Plasma Column in the Vacuum Arc
  Centrifuge}.
\newblock PhD thesis, UNIVERSITY OF SYDNEY, 2001.

\bibitem{Corr2009}
C.~S. Corr and R.~W. Boswell.
\newblock Nonlinear instability dynamics in a high-density, high-beta plasma.
\newblock {\em Physics of Plasmas}, 16(2), 2009.

\bibitem{Boswellthesis1970}
R.~W. Boswell.
\newblock {\em A Study of Waves in Gaseous Plasmas}.
\newblock PhD thesis, Flinders University of South Australia, 1970.

\bibitem{Kline2003}
J.~L. Kline, M.~M. Balkey, P.~A. Keiter, E.~E. Scime, A.~M. Keesee, X.~Sun,
  R.~Hardin, C.~Compton, R.~F. Boivin, and M.~W. Zintl.
\newblock Ion dynamics in helicon sources.
\newblock {\em Physics of Plasmas}, 10(5):2127--2135, 2003.

\bibitem{Scime2007}
E.~Scime, R.~Hardin, C.~Biloiu, A.~M. Keesee, and X.~Sun.
\newblock Flow, flow shear, and related profiles in helicon plasmas.
\newblock {\em Physics of Plasmas}, 14(4), 2007.

\bibitem{Plihon2005}
N.~Plihon, C.~S. Corr, and P.~Chabert.
\newblock Double layer formation in the expanding region of an inductively
  coupled electronegative plasma.
\newblock {\em Applied Physics Letters}, 86(9), 2005.

\bibitem{Dallaqua1998}
R.~S. Dallaqua, E.~Del~Bosco, R.~P. da~Silva, and S.~W. Simpson.
\newblock Langmuir probe measurements in a vacuum arc plasma centrifuge.
\newblock {\em Ieee Transactions on Plasma Science}, 26(3):1044--1051, 1998.

\bibitem{Hole2001}
M.~J. Hole and S.~W. Simpson.
\newblock Analytical description of a collisional plasma column in a vacuum arc
  centrifuge.
\newblock {\em Journal of Physics D-Applied Physics}, 34(20):3028--3035, 2001.

\bibitem{Lyman1962}
Lyman~Spitzer Jr.
\newblock {\em Physic of Fully Ionized Gases}.
\newblock John Wiley and Sons, 1962.

\bibitem{Shinohara2001}
S.~Shinohara, N.~Matsuoka, and S.~Matsuyama.
\newblock Establishment of strong velocity shear and plasma density profile
  modification with associated low frequency fluctuations.
\newblock {\em Physics of Plasmas}, 8(4):1154--1158, 2001.

\end{thebibliography}
\end{document}